\documentclass[usegraphicx,usenatbib,referee]{mn2e}

\begin{document}
\arraycolsep0.35mm                      % makes multiple equations look nice
\def\d{{\rm d}}\newcommand{\e}{{\rm e}}\def\b#1{{\bf#1}}
\newcommand{\be}{\begin{equation}}
\newcommand{\ee}{\end{equation}}
\newcommand{\kpc}{\,{\rm kpc}}
\newcommand{\kms}{\,{\rm km}~{\rm s}^{-1} }
\newcommand{\sm}{\,{\rm M}_{\sun}}
\newcommand{\mjup}{\,M_{\rm J}}
\newcommand{\au}{\,{\rm AU}}
\newcommand{\den}{\,{\rm M}_{\sun}\mbox{pc}^{-3}}
\newcommand{\dc}{D_{\rm c}}
\newcommand{\nc}{n_{\rm c}}
\newcommand{\ns}{n_{\rm s}}
\newcommand{\ds}{D_{\rm s}}
\newcommand{\vl}{v_{\rm l}}
\newcommand{\vs}{v_{\rm s}}
\newcommand{\ovs}{\overline{\b v}_{\rm s}}
\newcommand{\vo}{v_{\rm o}}
\newcommand{\vrot}{v_{\rm rot}}
\newcommand{\Gz}{\Gamma_0}

\title[Cold clouds as dark matter]{Observable consequences of
cold clouds as dark matter}
\author[Kerins, Binney \& Silk]{E. Kerins,$^{1,2}$ J. Binney$^2$
and J. Silk$^3$\\
$^1$Astrophysics Research Institute, Liverpool John Moores
 University, 12 Quays House, Birkenhead, Merseyside CH41 1LD.\\
$^2$Theoretical Physics, Department of Physics, University of Oxford,
 Keble Rd, Oxford OX1 3NP.\\
$^3$Nuclear \& Astrophysics Laboratory, Department of Physics,
University of Oxford, Keble Rd, Oxford OX1 3RH.}
\maketitle

\begin{abstract}
Cold, dense clouds of gas have been proposed as baryonic candidates
for the dark matter in Galactic haloes, and have also been invoked in
the Galactic disc as an explanation for the excess faint sub-mm
sources detected by SCUBA. Even if their dust-to-gas ratio is only a
small percentage of that in conventional gas clouds, these dense
systems would be opaque to visible radiation. This presents the
possibility of detecting them by looking for occultations of
background stars. We examine the possibility that the data sets of
microlensing experiments searching for massive compact halo objects
can also be used to search for occultation signatures by cold
clouds. We compute the rate and timescale distribution of stellar
transits by clouds in the Galactic disc and halo. We find that, for
cloud parameters typically advocated by theoretical models, thousands
of transit events should already exist within microlensing survey data
sets. We examine the seasonal modulation in the rate caused by the
Earth's orbital motion and find it provides an excellent probe of
whether detected clouds are of disc or halo origin.
\end{abstract}

\nokeywords
%\begin{keywords}
%Galaxy: halo -- Galaxy: kinematics and dynamics --
%microlensing -- dark matter
%\end{keywords}

\section{INTRODUCTION} \label{intro}

Over the last decade several workers have suggested that the Galaxy
may contain a large population of cold ($T \la 10$~K), dense, low-mass
clouds of molecular material. Several authors have argued that the
mass contained in such clouds could be responsible for keeping
galaxies' rotation curves flat beyond the edges of optical discs.
\citet{pfe94} argued that the clouds would be an integral part of a
fractal interstellar medium and be confined to a thin disc. By
contrast, \citet{dep95}, \citet{ger96}, \citet{dra98}, and
\citet{wal98} considered the case of cold clouds that are distributed
in an approximately spherical halo. De~Paolis et al, Draine and
Gerhard \& Silk argued for clouds of approximately solar mass, while
Walker \& Wardle pointed out that ionized winds from clouds with
masses around a Jupiter mass ($\mjup$) could explain the extreme scattering
events that are frequently observed in the light curves of pulsars
seen at low Galactic latitudes.

A major issue with these proposals that a significant mass is locked
up in very small clouds, is to understand why the clouds do not
collapse to form stars or degenerate objects such as brown
dwarfs. \citet{sci00} has argued that cosmic-ray heating may be able
to balance molecular cooling in a stable way for dust-free clouds,
whilst \citet{law01} has presented similar arguments that cosmic-ray
heating could balance cooling by dust emission in dusty clouds. Clouds
that are an integral part of the ISM would be expected to be dusty,
and radiate strongly at sub-mm wavelengths.
\citet{law01} has considered the possibility that a significant
fraction of the reddest SCUBA sources could lie in the Galaxy. He
concludes that for dusty clouds, masses between $\sim 10\mjup$ and
$\sim 10\sm$ can be ruled out, and that a population that contributes
significantly to the SCUBA counts will make a significant contribution
to the mass of the Galaxy only if they have masses $\sim M_{\rm
J}$. These objects would have to be confined to the Galactic plane.

The clouds would have temperatures $T\sim 10$~K, and if they were in
virial equilibrium, they would be $\sim 20$~AU in diameter and totally
opaque at optical wavelengths (and even have an optical depth of order
unity at 1~mm).  Their number density on the sky has to be comparable
to that, $\sim 1000$~deg$^{-2}$, of unidentified SCUBA sources, so
they would cover a fraction $\sim 10^{-6}$ of the sky. Consequently,
at any given time, of order one star in a million should be occulted
by one of these clouds.  It should be feasible to detect such
occultations in the databases that have been produced in searches for
microlensing events \citep[for a recent overview of microlensing
results see][and references therein]{ker01}.

Here we investigate the rate at which stars would be occulted by
opaque clouds.We argue in Section~\ref{detectability} that halo and
disc clouds are likely to have the characteristics required to produce
observable transits. In Section~\ref{theory} we derive general formulae for
the rate and timescale distribution of transit events and in
Section~\ref{predictions} we apply them to a specific
Galactic model. Currently available microlensing databases refer to a
handful of lines of sight. The largest body of data refers to lines of
sight in the general direction of the Galactic centre. Other extensive
data sets refer to sight lines towards the Magellanic
Clouds. Consequently, we concentrate on predictions for these
particular directions. We also investigate means by which halo and
disc cloud populations could be differentiated.

\section{Cloud characteristics and detectability} \label{detectability}

In order to occult background stars cold clouds must meet certain
criteria. Firstly, the clouds must be sufficiently diffuse that they
do not appreciably gravitationally lens the starlight. The case of
gravitational lensing by gas clouds has been studied previously by
\citet{hen95}.  A necessary condition to avoid the gravitational
lensing regime is that the cloud radius $R$ exceeds the Einstein
radius, leading to the requirement that
   \be
	R \gg 0.3\au \; \left( \frac{M}{10^{-3}\sm} \right)^{1/2}
	\left( \frac{D}{10\kpc} \right)^{1/2}, \label{radlim}
   \ee
where $M$ is the cloud mass and $D \equiv \dc (\ds - \dc)/\ds$, with
$\dc$ ($\ds$) the distance between the observer and cloud
(source).

The long-term stability of cold ($T \sim 10~K$) clouds is an open
issue and a serious theoretical problem for the scenario.  If
opaque clouds can be stabilized against collapse, their
size would be of the same order as their virial radius \citep[c.f.][]{law01}
	\be
	   R_{\rm vir} = 7\au \; \left( \frac{M}{10^{-3}\sm} \right)
	   \left( \frac{T}{10~K} \right)^{-1} \label{virrad}.
	\ee
$R \sim R_{\rm vir}$ would also satisfy the lensing condition in
equation~(\ref{radlim}) provided
	\be
	   M \gg 1.6 \times 10^{-6}\sm \; \left( \frac{T}{10~\mbox{K}}
	   \right)^2 \left( \frac{D}{10\kpc} \right) \label{masslim}.
	\ee
This condition is satisfied for almost all plausible clouds since,
regardless of their formation mechanism, hydrogenous clouds less massive
than $\sim 10^{-7}\sm$ evaporate away over a Hubble time \citep{deru92}.

Diffuse clouds will still refract, rather than occult, starlight if
they are transparent \citep{wal98,dra98}. How opaque might we expect the
clouds to be? The average hydrogen column density through a virialized cloud of
is
	\be
	   N_{\rm H} = 3.3 \times 10^{-29}~\mbox{m}^{-2} \; \left(
	   \frac{R}{7\au} \right)^{-1} \left( \frac{T}{10~K}
	   \right)^2.  \label{colden}
	\ee
Assuming typical dust properties, the visual extinction through a
cloud is therefore \citep[e.g.][]{bin98}
	\be
	   A_V = 1.7 \times 10^4 \, f \; \left( \frac{R}{7\au}
           \right)^{-1} \left( \frac{T}{10~K} \right)^2, \label{extin}
	\ee
where $f$ is the factor by which the cold cloud dust-to-gas ratio exceeds
that of typical Galactic molecular clouds. Equation~(\ref{extin})
implies that if cold clouds have even remotely the same dust abundance as
normal clouds then they easily fulfill the requirement that they be
opaque to starlight. In fact, for the clouds not to produce a detectable
occultation (which, conservatively, requires $A_V < 1$) they must
have a dust-to-gas ratio at least four orders of magnitude lower than
is measured for ordinary gas clouds, i.e. they must be essentially
dust free. 

We should expect disc clouds to have a similar dust-to-gas ratio as
the interstellar medium and therefore be completely opaque. However,
equation~(\ref{extin}) also implies strong limits on halo cloud
properties for them to escape detection by occultation. This can be
more clearly seen by considering how halo clouds may have interacted
with the interstellar medium over the lifetime $\tau_{\rm gal}$ of the
Galaxy.  Over this time one would expect a typical halo cloud to have
passed through the Galactic disc $N$ times, where
   \begin{eqnarray} \label{ncross}
	N & \sim & \sigma \tau_{\rm gal}/2 r_{\rm c} \nonumber \\
         & \sim & 160 \left( \frac{\sigma}{
        160 \kms} \right) \left( \frac{\tau_{\rm gal}}{10~\mbox{Gyr}} \right)
        \left( \frac{r_{\rm c}}{5\kpc} \right)^{-1},
   \end{eqnarray}
with $r_{\rm c}$ and $\sigma$ the halo core radius and
velocity dispersion. Neglecting accretion the mass swept up by these
clouds on each passage through an exponential disc is
   \begin{eqnarray} \label{msweep}
	M_{\rm sweep} & \sim & 2 \pi \varepsilon R^2 \rho H =
        4 \times 10^{-8}~\sm \, \varepsilon
	\left(\frac{R}{7~\mbox{AU}}\right)^2 \nonumber \\
        & & \hskip1.7cm \times \left(\frac{\rho}{0.03\den}\right)
        \left(\frac{H}{200~\mbox{pc}}\right),
   \end{eqnarray}
where $\rho$ is the typical density of gas in the plane of the disc,
$H$ the disc scale height and $\varepsilon$ the efficiency with which
the clouds retain swept up gas. Adopting the parameter values in
equations~(\ref{ncross}) and (\ref{msweep}) implies that the present
mass fraction of swept-up disc gas in a typical halo cloud is $NM_{\rm
sweep}/M \sim 0.007 \, \varepsilon$ for a $10^{-3}\sm$ cloud.This,
together with equation~(\ref{extin}), indicates that provided $\varepsilon >
0.008 f^{-1}$ the cloud dust-to-gas ratio should be large enough to
produce at least a one magnitude diminution in the light from background stars.

Lastly, we need to know how easily the flux change from transit events
can be detected.  An important consideration is source
resolvability. For Galactic microlensing experiments it is often the
case that more than one star occupies the seeing disc, leading to a
detection bias referred to as blending \citep{ala97}. The situation is
more extreme for experiments looking towards M31 where there are many
stars per detector pixel. Transit events could escape detection by
these surveys if the occulted star lies so close to other stars that
the effect of its occultation is masked by light from other
stars. Quantitatively, if a fraction $f_*$ of the flux in the seeing
disc comes from the target star, the flux at minimum will be a
fraction $(1-f_*)+f_*10^{-0.4A_V}$ of the flux prior to
occultation. Equating this fraction to $10^{-0.4\Delta M_T}$, where
$\Delta M_T$ is the smallest magnitude change a survey can measure, we find
that the event will be detected only if
   \be 
	f_* \ga \frac{1-10^{-0.4 \Delta M_{\rm T}}}{1-
   	10^{-0.4A_V}} \label{blend}
   \ee
For completely opaque clouds (large $A_V$) the requirement is simply $f_*
\ga \Delta M_{\rm T}$ for $\Delta M_{\rm T} \ll 1$. In the context of
gravitational microlensing searches, experiments to detect
microlensing events in our Galaxy typically adopt the very
conservative value $\Delta M_{\rm T} \sim 0.3$ as a means of excluding
variable stars. The M31 surveys typically use a much lower threshold
$\Delta M_{\rm T} \sim 0.02$. If similar thresholds are adopted for
the detection of transit events then they should be detectable within
our Galaxy provided blending effects are not too severe. It may even
be possible to detect transit events towards M31 provided the occulted
stars ordinarily contribute upwards of $2\%$ of the flux in a seeing disc.

\section{TRANSIT RATE AND TIMESCALE DISTRIBUTION} \label{theory}

From here onwards we assume that the cold clouds are opaque and exceed
their Einstein radius, so that they give rise to observable
occultation events. We calculate the rate at which occultations of a
given duration should be observed at any season of the year and along
any line of sight. Similar calculations of microlensing rates are
described by Griest (1991) and Kiraga \& Paczy\'nski (1994).

Let there be $n(\b v)\d^2\b v$ spherical clouds of radius $R$ per unit
volume and with velocities in the plane of the sky in $\d^2\b v$. Then in
a sheet of thickness $\d \dc$ that extends perpendicular to the line of
sight, the rate at which clouds pass a given background star with impact
parameters in the range $(b,b+\d b)$ is
   \be
	\d^4 \Gamma = n(\b v)\d^2\b v\,\d \dc\,2v_\perp\d b,
   \ee
 where $v_\perp=|\b v|$. In an obvious notation we have $\d^2\b v=v_\perp\d
v_\perp\d\theta$, so
   \be \label{dr1}
	\d^4 \Gamma = 2n(\b v) v_\perp^2\d v_\perp\,\d\theta\,\d \dc\,\d b.
   \ee
An impact with  $b<R$ causes an occultation of duration
 \be\label{defst}
t=2\sqrt{R^2-b^2}/v_\perp,
 \ee
 so we may obtain the distribution over durations by eliminating $b$
in favour of $t$. One finds
   \be\label{dt1}
	{\d^4 \Gamma\over \d \dc \d t} = 
{n(\b v) v_\perp^3\d v_\perp\,\d\theta\,
\over\sqrt{(2R/v_\perp t)^2 - 1} }
   \ee
This equation gives the occultation rate of a single source star. To
get the rate that can be measured observationally, we have to average
it over all detectable source stars \citep[c.f.][]{kir94}. If the
probability that a source star at distance $\ds$ has its velocity on the sky in
$\d^2\b \vs$ is $p(\b \vs)\d^2\b \vs$, then the required average  of equation
(\ref{dt1}) is
 \be\label{dt2}
	{\d^2 \Gamma\over \d \dc \d t} =
\int\d^2\b \vs\,  p(\b \vs)
\int_0^{2R/t}\hskip-15pt \frac{\d v_\perp\,v_\perp^3} {
        \sqrt{(2R/v_\perp t)^2 - 1} } \int\d\theta\, n(\b v).
\ee
$v_\perp$ is the magnitude of the cloud's transverse velocity with
respect to its local point on the line of sight from the source to the
Sun; consequently it is a function of $\b \vs$. The argument $\b v$ of
the velocity distribution $n(\b v)$ must relate to a fixed frame.  In
the frame in which the Earth is at rest
 \be\label{givesvp}
v_\perp=\left|\b v-{\dc\over \ds}\b \vs\right|,
 \ee
 where $D_s$ is the distance to the source star,
so, for fixed $v_\perp$, $\b v$ is a function of $\b v_s$ and to make
further progress analytically, it is necessary to assume that both $\b
v$ and $\b \vs$ have Gaussian distributions. Since we are working in
the Earth's rest frame, these Gaussians will be centred on non-zero
velocities $\overline{\b v}$ and $\overline{\b v}_s$,
respectively. Consequently we have
 \be\label{prodPN}
p(\b \vs)n(\b v)={n_c\over(2\pi\sigma_s\sigma)^2}\exp\left[-{(\b \vs-\ovs)^2\over2\sigma_s^2}-{(\b v-\overline{\b v})^2\over2\sigma^2}\right],
\ee
where $n_c$ is the volume density of clouds.
Now $\theta$ is the direction
of $\b v-(\dc/\ds)\b \vs$, so with (\ref{givesvp}) we have
 \begin{eqnarray}\label{argexp}
(\b v-\overline{\b v})^2&=&\left|
\b v-{\dc\over \ds}\b \vs+{\dc\over \ds}\b \vs-\overline{\b v}
\right|^2\nonumber\\
&=&v_\perp^2+2v_\perp \vs\cos\theta
+\vs^2,
\end{eqnarray}
 where
\be\label{defsVs}
\vs\equiv \left|{\dc\over \ds}\b \vs-\overline{\b v}\right|.
\ee
 When we insert (\ref{argexp})  into $n(\b v)$ and integrate over
 $\theta$ we find
\be
\int\d\theta\, n(\b
v)={n_c\over\sigma^2}\exp\left[-{v_\perp^2+\vs^2\over2\sigma^2}\right]I_0
\left({v_\perp \vs\over\sigma^2}\right),
\ee
 where $I_0$ is the usual modified Bessel function.
Inserting this result in (\ref{dt2}) we find
\begin{eqnarray}\label{d2gamma1}
{\d^2 \Gamma\over\d \dc\d t}& =&{\nc\over2\pi\sigma_s^2\sigma^2} 
\int_0^{2R/t}\hskip-.5cm{v_\perp^3
\,\d v_\perp \over\sqrt{(2R/v_\perp t)^2 - 1} } 
\nonumber\\
&& \hskip-1.cm \times\int\d^2\b \vs\exp\left[-{(\b \vs-\ovs)^2\over2\sigma_s^2}
-{v_\perp^2+\vs^2\over2\sigma^2}\right]I_0\left({v_\perp
\vs\over\sigma^2}\right).
 \end{eqnarray}
This result simplifies significantly if the sources have negligible 
random motions
(i.e. $\sigma_s\dc/\ds\ll\sigma$). In this case we can approximate the
Gaussian distribution over $\b \vs$ by $\delta^2(\b \vs-\ovs)$, giving
 \begin{eqnarray}\label{d2gamma2}
{\d^2 \Gamma\over\d \dc\d t} &\simeq& {\nc\over\sigma^2}
\int_0^{2R/t}\hskip-.5cm{\d v_\perp\,v_\perp^3\over
\sqrt{(2R/v_\perp t)^2 - 1} } \nonumber\\
&& \quad \quad \quad \quad \quad \times\exp\left[
-{v_\perp^2+\vs^2\over2\sigma^2}\right]I_0\left({v_\perp
\vs\over\sigma^2}\right).
 \end{eqnarray}
 with $\ovs$ replacing $\b \vs$ in (\ref{defsVs}).
The rate of occultations integrated over all durations$\int\d
t\,\d^2\Gamma/\d \dc\d t$ is readily obtained from either
(\ref{d2gamma1}) or (\ref{d2gamma2}):
\begin{eqnarray}\label{totalr}
{\d\Gamma\over\d \dc}
& =&{\nc R\over\pi\sigma_s^2\sigma^2}
\int\d v_\perp\,v_\perp^2\int\d^2\b \vs
\nonumber\\
&& \quad \times \exp\left[-{(\b \vs-\ovs)^2\over2\sigma_s^2}
-{v_\perp^2+\vs^2\over2\sigma^2}\right]I_0\left({v_\perp
\vs\over\sigma^2}\right)\\
&\simeq&{2\nc R\over\sigma^2}
\int\d v_\perp\,v_\perp^2 \exp\left[
-{v_\perp^2+\vs^2\over2\sigma^2}\right]I_0\left({v_\perp
\vs\over\sigma^2}\right).\nonumber
 \end{eqnarray}
Since the central velocities $\ovs$ and $\overline\b v$ are
relative to the Earth, equations (\ref{d2gamma1}) to (\ref{totalr}) yield
occultation rates that depend on observing season.

From equation~(\ref{totalr}) $\Gamma \propto R/M$
(because $\nc \propto \rho/M$), so for virialised clouds
equation~(\ref{virrad}) indicates that the transit rate is controlled
simply by the cloud temperature $T$, i.e. $\Gamma \propto T^{-1}$,
independent of the scale size of the cloud. However, for fixed
dimensionless impact parameter $b/R$ the transit time $t \propto R$
from equation~(\ref{defst}) so $\d \Gamma/\d t \propto R^{-1}T^{-1}$
when integrated over all impact parameters $0 \leq b/R \leq 1$.

In some applications the sources are widely distributed in distance
$\ds$.  In this case we follow Kiraga \& Paczy\'nski (1994) in
evaluating weighted averages over $\ds$ of the rates (\ref{d2gamma1})
to (\ref{totalr}), with the weight factor taken to be $n_{\rm s}
\ds^{2+2\beta}$. Here $n_{\rm s}$ is the source volume number density
and $\beta$ parameterizes the effects of the luminosity function upon
the distance distribution of sources.  In our calculations we adopt
$\beta = -2$ for $\ds > 100\,$pc and $\beta = -1$ at smaller
distances.

\section{MODEL PREDICTIONS} \label{predictions}

We explore two  locations for potential cloud populations, the
Galactic disc and halo, and we compute the rate of occultations for lines
of sight to three targets: the Large Magellanic Cloud (LMC) at
Galactic coordinates $(l, b) = (280\degr, -33\degr)$, the Small
Magellanic Cloud (SMC) at $(l, b) = (303\degr, -44\degr)$ and Baade's
Window towards the Galactic Centre (GC) at $(l, b) = (1\degr,
-4\degr)$. These are all regions which have been intensively monitored
by several experiments looking for microlensing events. As discussed
in Section~\ref{detectability}, a fourth possibility is the Andromeda
Galaxy (M31), though in that case the formulae of the previous Section
must be modified to take account of the fact that the sources in M31
are mostly unresolved. We therefore confine our attention in the
remainder of this paper to the GC, LMC and SMC only. The LMC and SMC
source stars are assumed to lie at fixed distances of $50\kpc$ and
$60\kpc$, respectively, whilst background sources towards Baade's
Window are assumed to be distributed throughout the Galactic disc and
bar.

Since we are not concerned with the precise details of the halo
structure, it is sufficient for our purposes to model the distribution
of halo clouds as a simple softened isothermal sphere with a core
radius of $5\kpc$ and a local density of $0.01\den$. The halo velocity
distribution is taken to be Gaussian with $\sigma = 156\kms$. For both
clouds and sources in the Galactic disc we model their density
distribution as a sech-squared profile with a scale length of
$2.5\kpc$ and scale height of 190~pc. We assume the clouds contribute up
to a third of the disc density and so have a local density of
$0.03\den$. Strong local kinematical and dynamical constraints rule
out the existence of a more significant population \citep{cre98}. We
adopt a velocity dispersion of $\sigma = 25\kms$ and a rotation speed
$\vrot = 220\kms$ for both disc clouds and sources, independent of
radius.

For calculations towards the LMC and SMC we neglect random stellar
motions in the clouds themselves though we do include the bulk motion
of the Clouds $(u, v, w) = (60, -155, 146)\kms$ \citep{jon94}, where
the $(u,v,w)$ coordinate system is centred on the Sun and the velocity
components point, respectively, towards the Galactic Centre, the
direction of rotation, and the North Galactic Pole.
Equations~(\ref{d2gamma2}) and (\ref{totalr}) are therefore sufficient
to compute the halo cloud timescale distribution and rate towards
these targets. For bulge
sources we adopt the bar model of
\citet{bis97} with exponential scale length $a_{\rm m} = 3.82\kpc$,
power-law scale length $a_0 = 100$~pc, axis ratios $\eta = \zeta =
0.3$ and bar angle $\phi_0 = 25\degr$. We assume a bar pattern speed
of $70\kms\kpc^{-1}$ and a velocity dispersion $\sigma =
100\kms$. Whilst our disc model is simpler than the double-exponential
profile of \citet{bis97} its scale length is the same and the chosen
scale height corresponds to the weighted mean of the scale heights in
their model. Our overall disc-bulge mass normalization is therefore in
good agreement.

The motion of the observer is modeled as the sum of three distinct
components: the motion of the LSR $(u,v,w) = (0, \vrot, 0)\kms$; the
Sun's motion in the direction of the Solar Apex $(u,v,w) = (10, 5.2,
7.2)\kms$ \citep{bin98}; and the motion of the Earth around the Sun,
where for simplicity we adopt a circular orbit with $v_{\oplus} =
30\kms$.

\subsection{Rate and timescale distribution}

\begin{figure}
\includegraphics[scale=0.35,trim= 0 0 -25 0,angle=270]{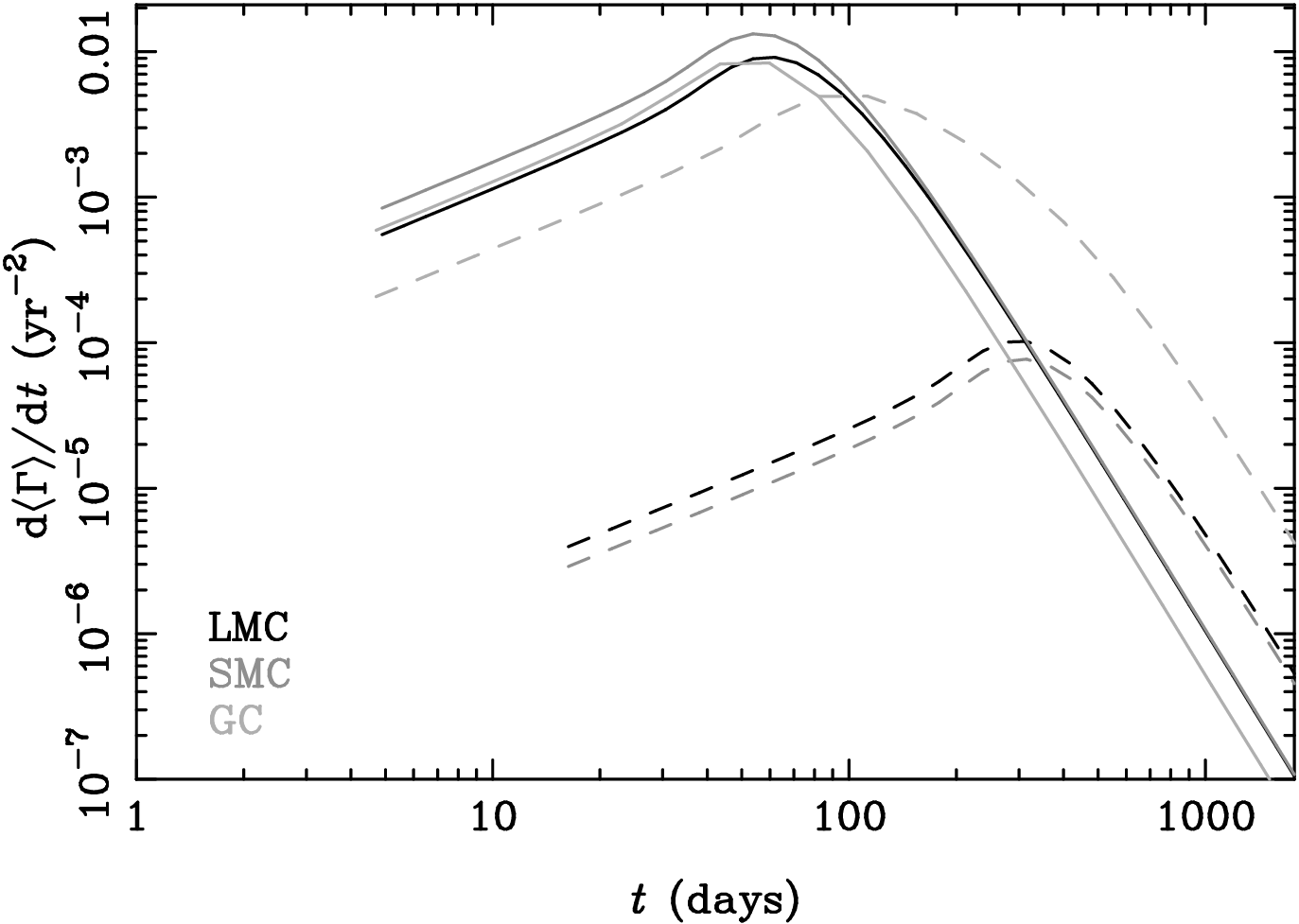}
\includegraphics[scale=0.35,angle=270]{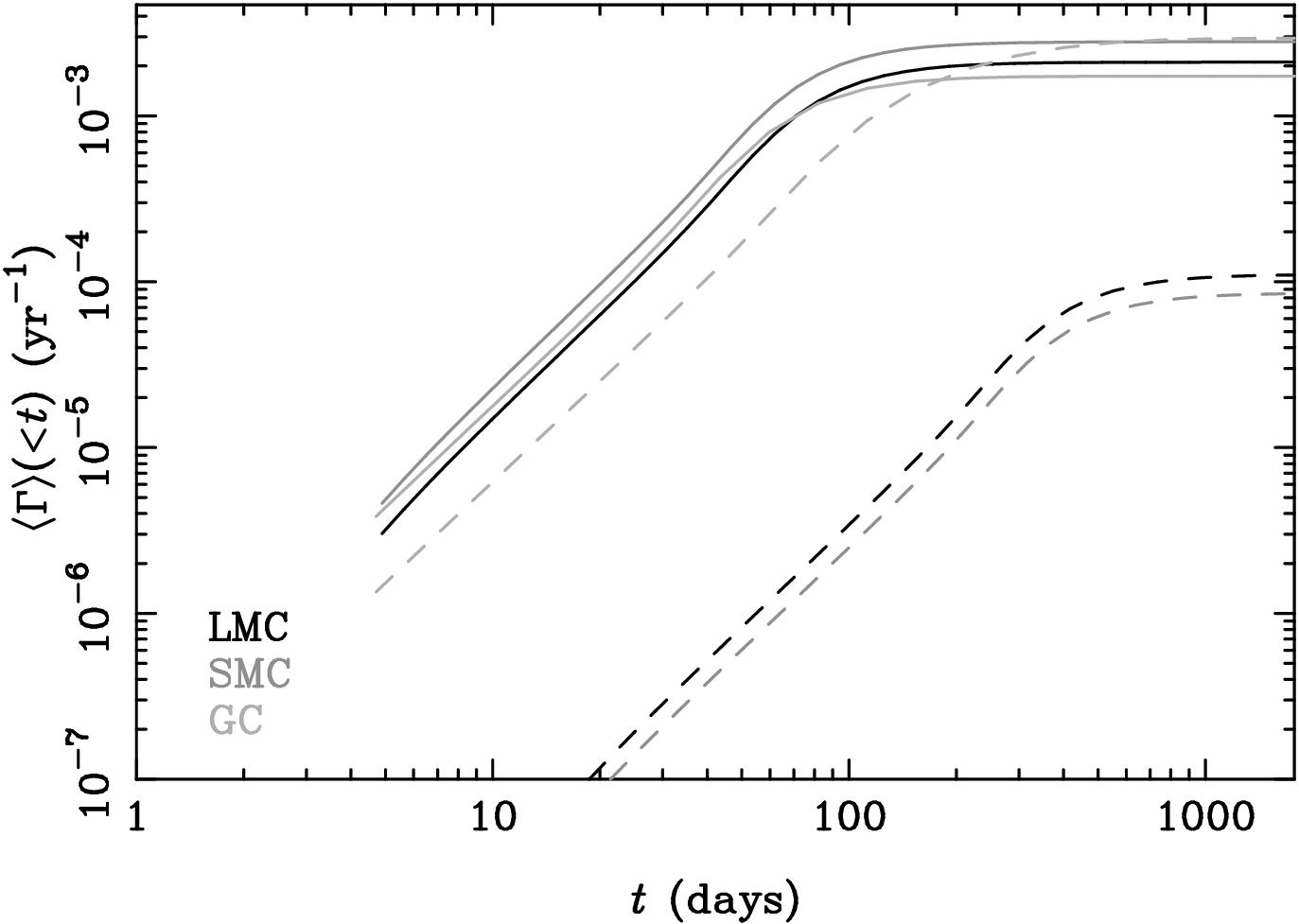}
\caption{Seasonally-averaged differential (top) and cumulative
(bottom) cloud transit timescale distributions for halo clouds, (solid
lines) and disc clouds (dashed lines). The distributions are shown for
three lines of sight: LMC (black); SMC (dark grey); GC
(light grey). The figures assume the clouds have a radius $R =
7$~AU and virial temperature $T = 10~\mbox{K}$. The scaling for other
values is $\d\Gamma/\d t \propto R^{-1}T^{-1}$ and $t \propto R$.}
\label{f1}
\end{figure}

Fig.~\ref{f1} shows the predicted distributions of occultation time
for both halo and disc clouds towards the LMC, SMC and GC. The
distributions are seasonal averages of the timescale
distribution. We have adopted a radius $R = 7$~AU and virial
temperature $T = 10$~K for both cloud populations, corresponding to a
cloud mass of $M = 10^{-3}\sm$ from equation~(\ref{virrad}). The top panel
of Fig.~\ref{f1} shows the differential distribution, whilst the
lower panel gives the cumulative rate. For these parameters we see
that the halo cloud distributions (solid lines) are peaked at shorter
durations than for disc clouds (dashed lines), with the peak occurring
around 60 days for all three target directions. For disc clouds the
peak occurs at around 100 days in the direction of the GC or around
300 days towards the LMC and SMC.

For fixed cloud temperature the most detectable signals comes from
disc clouds observed towards the GC and halo clouds towards the SMC,
both with total rates of $\sim 3000$ events/yr per million stars. Halo
clouds towards the SMC have the largest rate for transit durations
below 500 days. Halo clouds observed towards the LMC and GC have
smaller, but comparable, rates. These large rates in comparison with
microlensing is due to the size of the clouds. For $10^{-3}\sm$ clouds
their virial radius is 1.5--2 orders of magnitude larger than the
typical Einstein radius of compact objects of the same mass, so the
rate is correspondingly higher. Both the timescales and the expected
rate are easily within range of current microlensing
experiments. Towards the LMC and SMC the disc cloud transit rate is
lower by more than an order of magnitude at $\sim 100$ events/yr per
million stars. Even when detection efficiencies have been taken into
account, microlensing data sets should contain hundreds or even
thousands of transit events, so it is clear that much of the parameter
space for both halo and disc cloud populations can be probed.

\subsection{Discriminating between disc and halo clouds} \label{modulation}

Can we discriminate between halo and disc cloud populations?  From
Fig.~\ref{f1} it is apparent that a rate of $\ga 1000$ events/yr per
million sources towards the LMC and SMC would argue strongly for a
halo origin.  Another obvious diagnostic is the transit
duration. Fig.~\ref{f1} indicates that we should expect few halo
cloud events with transit durations above 100 days, whilst more than
three-quarters of the rate of disc clouds towards the GC comes from
events with durations above 100 days. Of course larger halo clouds can
produce longer transits so this is not a robust test.

\begin{figure}
\includegraphics[scale=0.35,angle=270]{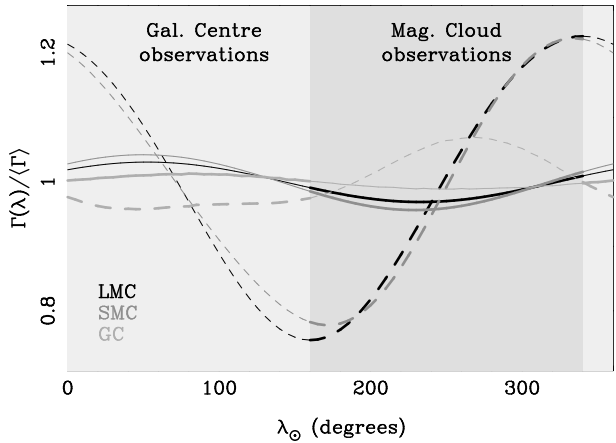}
\caption{Seasonal modulation of the occultation rate. Displayed as a
function of Solar Ecliptic longitude, $\lambda_{\sun}$, is the
instantaneous rate, $\Gamma$, expressed as a ratio of the
seasonally-averaged rate, $\langle \Gamma \rangle$. Line coding is as
for Fig.~\ref{f1}. The approximate observing season for the LMC and SMC is
shown by the dark grey shaded region whilst the GC monitoring
period is shown in light grey. The modulation curves are shown as
thick lines during the relevant observing season and as thin lines
otherwise.}
\label{f2}
\end{figure}

A potentially cleaner test is the seasonal modulation in the
observed rate due the Earth's
motion around the Sun. The modulation is most pronounced for nearby clouds, 
which will typically be disc clouds.
Fig.~\ref{f2} shows the modulation in the instantaneous transit rate
$\Gamma$, as a ratio of the seasonally-averaged rate $\langle \Gamma
\rangle$. The modulation is plotted as a function of Solar Ecliptic
Longitude, $\lambda_{\sun}$, where $\lambda_{\sun} = 0$ at the Vernal
Equinox which occurs around March~21st. The dark grey region of the
figure, spanning from September to March, indicates the six-month
period when microlensing experiments monitor the Magellanic
Clouds. For the rest of the year the GC is observed, as shown by the
light grey region. Consequently, only half of the modulation period is
accessible for any of the targets, so the modulation curves
are shown as bold lines whilst the relevant target is observed and as
thin lines when not.

The largest modulation occurs for disc clouds observed towards the LMC
and SMC. During the LMC/SMC observing season the rate swings from 0.8
to 1.2 times the seasonal average, a modulation of $20\%$. Assuming
$100\%$ detection efficiency, a $20\%$ modulation could be detected at
the $3\, \sigma$ level after four seasons of data for a million
sources. Current experiments are monitoring closer to $10^7$ sources
towards the LMC, though their detection efficiency to transits will be
somewhat less than $100\%$. It's nonetheless clear that this
modulation should be detectable already.  Halo clouds show only a
small (few percent) variation in their rate towards the LMC and SMC,
however their transit rate is much larger than for disc clouds. A sample
upwards of 30000 transit events is required to detect a modulation of
$2\%$ at greater than $3\, \sigma$ significance. Experiments
monitoring $10^7$ sources would amass such a sample within four years,
so even this small modulation should be evident within existing data
sets. Towards the GC the modulation of the disc and halo cloud rate is
similarly small but the rate is comparably large, so again the
modulation should be apparent. In the case of halo clouds,
observations towards the GC should show a maximum rate during
mid-season, whilst disc clouds should exhibit a minimum. The
modulation signal should therefore provide a good test for the origin
of detected events.

\section{DISCUSSION}

Cold opaque clouds can be detected by searching 
data sets already compiled by groups  searching  for
gravitational microlensing events. Rather than the flux excess that
is the signature of microlensing, cold cloud events are characterized
by a transient dimming of background sources. Like microlensing, this
signal should be non-periodic and one should expect events to involve
a representative population of background source stars. In selecting
microlensing events one problem is to reject variable stars
which may mimic a microlensing signal. The equivalent
``background'' for cloud transit events is potentially much smaller
since there are fewer classes of objects which involve a significant
dimming of  flux. One such class, eclipsing binaries, is
unlikely to present much of a problem. For short-period systems their
true nature would be evident over the lifetime of the surveys. Binary
systems with longer periods for which only one flux dropout
is detected, should statistically comprise more massive stars, so we
should expect the source stars of these ``events'' to be
unrepresentative of the target stellar population as a whole, contrary
to expectation for true cloud transit events. It is also likely that
the light-curve of cloud transit events would have a form which
generally would be inconsistent with that of eclipsing binaries.

The previous section demonstrates that the transit rate and timescales
of cold clouds in the disc and halo is well within the range of
detectability if they constitute a significant population and are of a
Jupiter mass scale. Whilst the total rate depends only on the
temperature of the clouds, virialized clouds more massive than $\sim
0.05\sm$ would have transit durations exceeding the present baseline
of microlensing experiments.

The seasonal modulation of the rate provides a promising method to
distinguish whether the clouds are in the disc or the halo. There are
already strong arguments from the consideration of sub-mm sources
\citep{law01} that opaque cold clouds do not contribute much to the
halo dark matter budget, though these arguments are sensitive to the
precise dust properties of the clouds. Existing microlensing data sets
represent a significant corpus of data which can provide an
independent line of approach. Though originally conceived as a probe
only of compact objects, microlensing surveys may also turn out to be
one the most sensitive probes of diffuse cold clouds.

%\section*{acknowledgments}

\end{document}